\documentclass{article}
\usepackage{graphicx}
\def\le{\underline{<}}
\begin{document}

\title{Vitrification of a monatomic 2D simple liquid}

%\articletype{Research Article}

\author{Tomoko Mizuguchi and Takashi Odagaki\\
Department of Physics, Kyushu University, Fukuoka 812-8581 Japan% etc
%\email{t.odagaki@cmt.phys.kyushu-u.ac.jp}
}                     % Do not remove
%\institute{Department of Physics, Kyushu University, Fukuoka 812-8581 Japan}
\date{ }
\maketitle
\abstract{
A monatomic simple liquid in two dimensions, where atoms interact isotropically
through the Lennard-Jones-Gauss potential [M. Engel and H.-R. Trebin, Phys. Rev. Lett. 
98, 225505 (2007)], is vitrified by the use of a rapid 
cooling 
technique in a molecular dynamics simulation.
Transformation to a crystalline state is investigated at various temperatures
and the time-temperature-transformation (TTT) curve is determined. It is found that
the transformation time to a crystalline state is the shortest at a 
temerature 14\% below the melting temperature $T_m$ and that at temperatures below 
$T_v \equiv 0.6 T_m$ the transformation time is much longer than the available CPU 
time. This indicates that a long-lived glassy state is realized for $T \le T_v$.
 %end of abstract
%
%\keywords{vitrification \*\ Lennard-Jones-Gauss potential \*\ crystallization}
%\pacs{64.70.P-,05.20.Jj,68.18.Jk}
% end of PACS codes
%
%

%
\section{Introduction}
\label{sec:1}
Glassy materials have been utilized by human beings since the stone age,
but still the transition from a liquid to a glass is far from 
fully understood on the basis of physics. Difficulty of the 
understanding lies in the fact that there are two effects in the vitrification 
process; one is the chemical order due to entangled networks or many 
constituents, and the other is the topological order due to 
geometrical jamming.

In fact most of the molecular dynamic (MD) simulations used to 
observe glass transition have been carried out for liquids with more than two 
components in order to avoid crystallization\cite{hiwatari,kob}.
It is known that simple metallic liquids can be vitirified by 
rapid quenching, but complicated constituents are needed to obtain 
a long-lived glassy state\cite{mrs}. Colloidal suspensions can be 
vitrified, but it is known that the system is eventually crystallized\cite{megen}.
Since most theoretical works\cite{mct,fel,replica} assume a 
simple system, comparison between theory and experiments is rather difficult.
Therefore, it is very desirable that we can devise a potential
in such a way that atoms interacting with it can be vitrified even if the system 
consists of one component. Such a potential allows us to separate the contribution 
of the topological order from the chemical one in the glass formation.

For the purpose of producing a long-lived glassy state of a monatomic simple liquid, 
we employ the Lennard-Jones-Gauss (LJG) potential\cite{engel}. In a previous 
report\cite{mizuguchi}, we showed that the LJG system can be vitrified in two 
dimensions and the glassy state at low temperatures is stable for a fairly long time 
in spite of a simple monatomic potential. In addition, for 3D systems the 
glass-forming ability of the LJG system has been tested and discussed\cite{hoang}. 
It is important to note that since any system can be brought into an amorphous state 
by rapid cooling, one has to investigate if the amorphous state is stable for a long 
time and shows a glass transition when heated.

In this paper, we investigate the aging-induced crystallization for the monatomic 
glass-forming system with MD simulations. In general simulations, the model system 
is chosen to favour the clarification of dynamical relaxations under supercooling, 
while the crystallization is highly suppressed. Therefore, correct 
discriptions of the relaxation process of the glass-forming liquids are still not 
really understood. We focus attention on this point and proceed 
with the study of the crystallization process 
within the model reported here which have a long relaxation time.
In Sec.2, we explain our model and methods of MD simulation in detail. Results of 
the simulation are presented in Sec.3 and we discuss the results in Sec. 4.

\section{Model and molecular dynamics simulation}
\label{sec:2}
We consider a monatomic system in two dimensions
where atoms interact with each other
isotropically via the Lennard-Jones-Gauss (LJG) potential
\begin{equation}
V(r)=\epsilon_0 \left\{ \left( \frac{r_0}{r} \right)^{12}-
2\left( \frac{r_0}{r} \right)^{6} \right\}-
\epsilon \exp \left( -\frac{(r-r_G)^{2}}{2\sigma ^{2}} \right).
\label{ljgpot}
\end{equation}
The LJG potential is of a double-well type for most values of
the parameters, with the second well of depth $\epsilon$ and width $\sigma$
at position $r_G$. 
We note that the effective potential, similar to the LJG
type, exists in metals\cite{goukin} and has been proposed
as a model potential to stabilize certain structures\cite{rechtsman}.
\begin{figure*}[hbtp]
\centering
%\resizebox{0.5\textwidth}{!}{%
\includegraphics[width=0.5\textwidth]{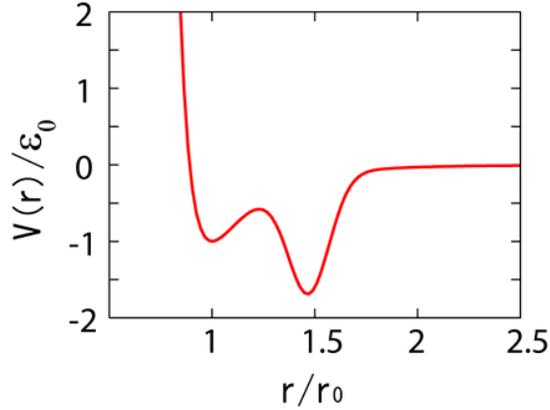}
%}
\caption{The LJG potential (\ref{ljgpot}) for 
$r_G=1.47r_0$, $\epsilon=1.5\epsilon_0$, $\sigma^2=0.02r_0^2$.}
\label{ljg}
\end{figure*}%

In this paper, we fix the parameters to
$r_G=1.47r_0, \epsilon=1.5\epsilon_0, \sigma^2=0.02r_0^2$
and investigate if stable glassy states can be obtained by quenching.
Figure 1 shows the LJG potential with these fixed parameters.
This set of parameters produces the pentagon-triangle phase
in the ground state, where nearest neighbour links form
tiling of pentagons, squares and triangles\cite{engel}.

In the following MD simulation, we set the unit of length, energy and time
as $r_0$, $\epsilon_0$, $\tau \equiv ( mr_0^2/\epsilon_0)^{1/2}$, respectively,
and the scaled temperature is denoted by $T^* \equiv k_BT/\epsilon_0$.
The number of atoms in the system is $N=1,024$ and the free boundary is used.
We employ the Leap-Frog algorithm for the MD simulation with time step $0.01\tau$.

We first prepared an equilibrium liquid state at $T^* = 0.4$ above the melting
temperature $T^*_m\simeq 0.34$. Then, we quenched the system to zero temperature 
instantaneously by removing the kinetic energy. We employ the 
instantaneous cooling in order to avoid any nucleation 
during the cooling process which may hinder proper comparison of 
the crystallization process. After the quenching, we subjected the 
system to a heat reservoir and allowed it to relax at 
this temperature. The simulation cell is made 
large enough so that the system feels zero pressure at all times. 
Under this free boundary condition, particles on the surface can easily move. 
Therefore, rearrangement of particles can frequently occur, compared 
with periodic boundary conditions or a system confined in a small 
space\cite{nishio}, and cystallization will proceed more quickly and smoothly. We 
observed the structural transformation in real space as a function of observation 
time.

\section{Results}
\label{sec:3}
In order to determine the crystallization time, we focus on the formation of three 
characteristic tiling structures, PPPT, PPP and PPTS, which consist of pentagons, 
squares and triangles shown in Fig. \ref{tiling} (a).
Figure \ref{tiling} (b) shows the time dependence of the number of
atoms whose surroundings can be identified as the three characteristic 
tiling structures.
\begin{figure*}[htbp]
\centering
%\resizebox{0.5\textwidth}{!}{%
\includegraphics[width=0.5\textwidth]{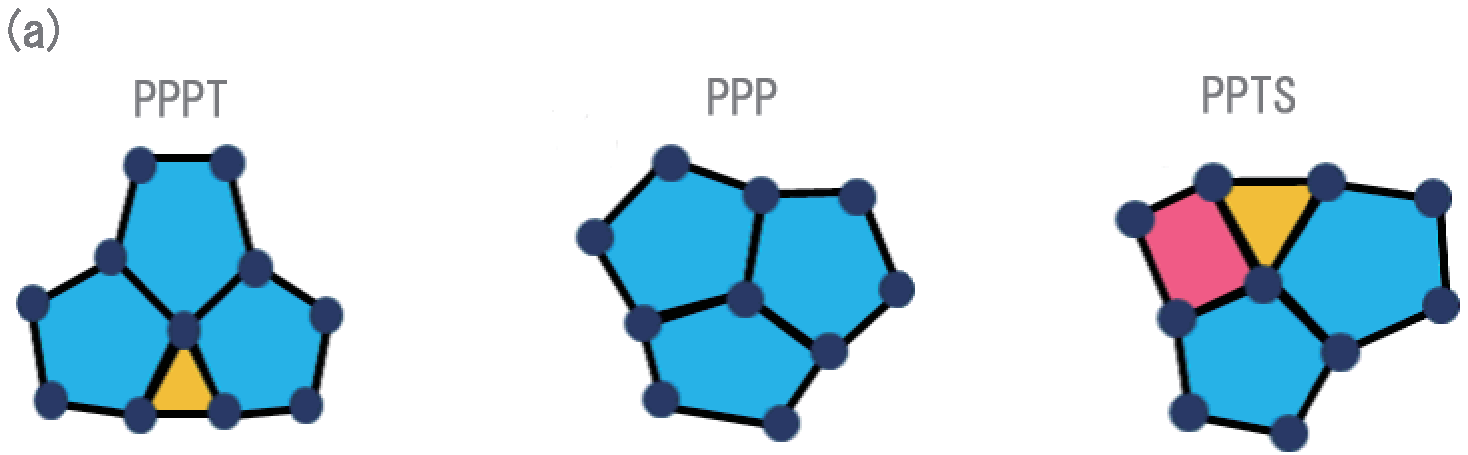}
%}

%\resizebox{0.5\textwidth}{!}{%
\includegraphics[width=0.5\textwidth]{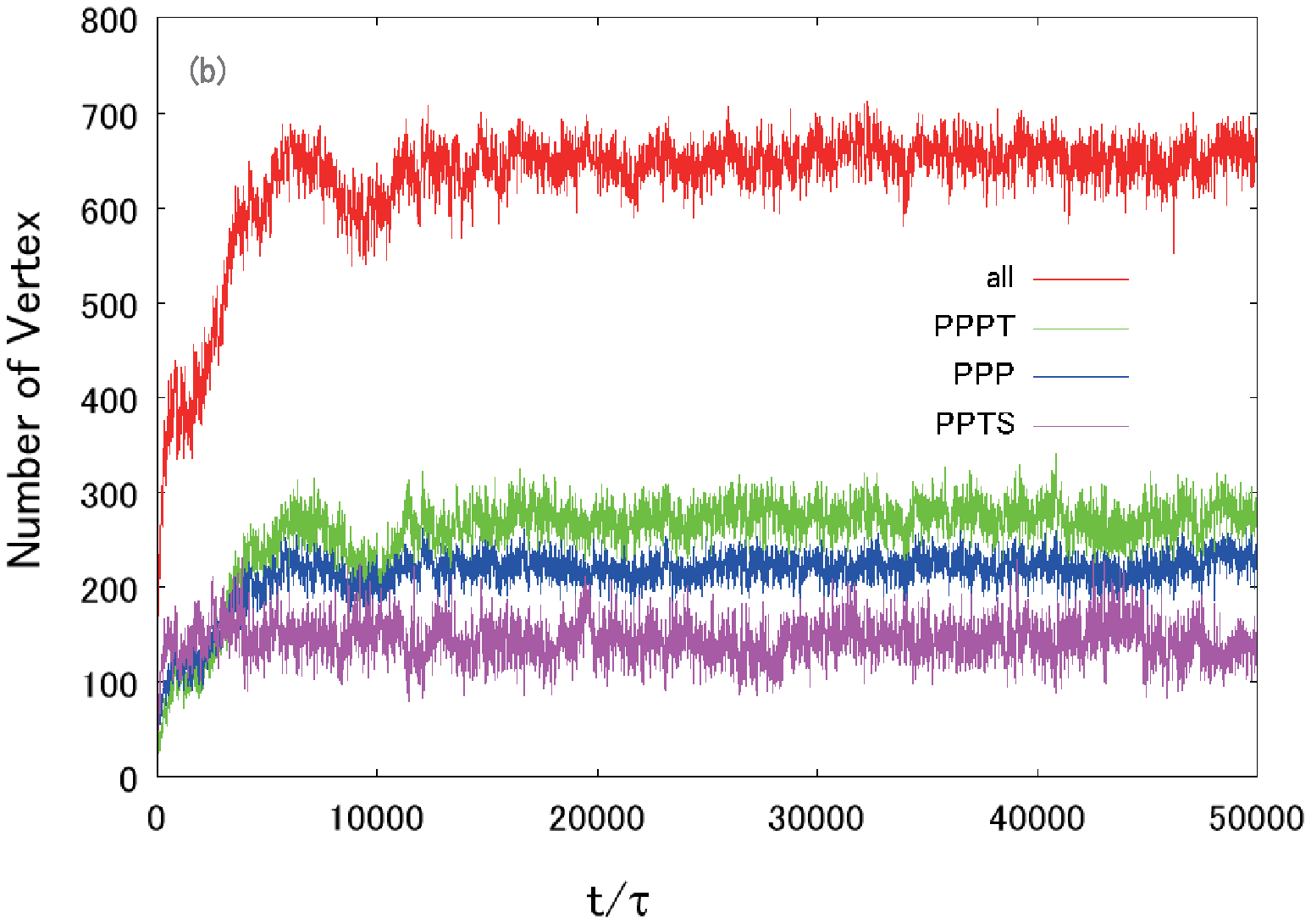}
%}
\caption{(a) Three charateristic tiling structures.
(b) The time dependence of the number of characteristic tiling
 structures. The curve (all) represents the number of atoms surrounded 
by any of the pentagon, triangle or square tiles. The curves (PPPT), (PPP) 
and (PPTS) correspond to the structures shown in (a).}
\label{tiling}
\end{figure*}%
The curve (all) shows the number of particles surrounded by any of the pentagons, 
triangles or squares. This value is related to the potential energy and becomes 
constant when the system reaches a crystalline state. At relatively high 
temperatures, the transition from the supercooled liquid to a crystal is rather 
sharp and we can determine the crystallization time without 
ambiguity. As  temperature is reduced, 
the transformation tends to be less sharp, and  we determined the transition time 
with a diffraction pattern as well as the evolution of tilings.
Figure \ref{TTTdiagram} shows the crystallization time as a function of temperature
which is known as a time-temperature-transformation (TTT) diagram.
This diagram indicates that the crystallization time becomes the shortest
at $\sim 0.88 T_m$ and it is longer than $10^7 \tau$ at $0.5 T_m$.
If one uses the value of $m$, $r_0$ and $\epsilon_0$ relevant for Ar,
this time corresponds to $10$ micro seconds.

For comparison, the results for a monatomic Lennard-Jones(LJ) system are also plotted
in Fig. \ref{TTTdiagram}, where the crystallization time is about $10^3$ MD steps. 
Consequently, the LJG system has a much longer crystallization time 
than the LJ system, and we can clearly see the temperature dependence of the TTT 
diagram in the LJG system. It shows a typical nose shape, which has 
been observed by experiments for various glass forming 
materials\cite{mrs,TTT1,TTT2}. A similar TTT diagram is also found by MD simulations 
with some empirical potentials for metal\cite{TTT_Fe,TTT_Ni}. Our 
system has a much longer crystallization time than that reported in 
these papers in spite of the simpler shape of our interatomic 
potential, if we use the parameters relevant for Ar.

\begin{figure*}[htbp]
\centering
%\resizebox{0.5\textwidth}{!}{%
\includegraphics[width=0.5\textwidth]{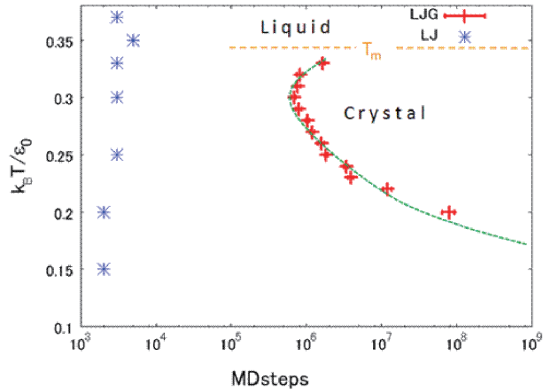}
%}
\caption{The time-temperature-transformation diagram for 
LJG and LJ liquids.}
\label{TTTdiagram}
\end{figure*}%

In Fig. \ref{crystallization}, we show the atomic configurations at two different temperatures,
$T^* = 0.3$ and $0.15$, and two different observation times $t/\tau = 3\times 10^2$ and
$10^5$. We can clearly see that at $T^* = 0.3$ crystallization has 
taken place in this time period and that at $T^* = 0.15$ the system 
keeps the amorphous structure.

\begin{figure*}[htbp]
\centering
%\resizebox{0.6\textwidth}{!}{%
\includegraphics[width=0.5\textwidth]{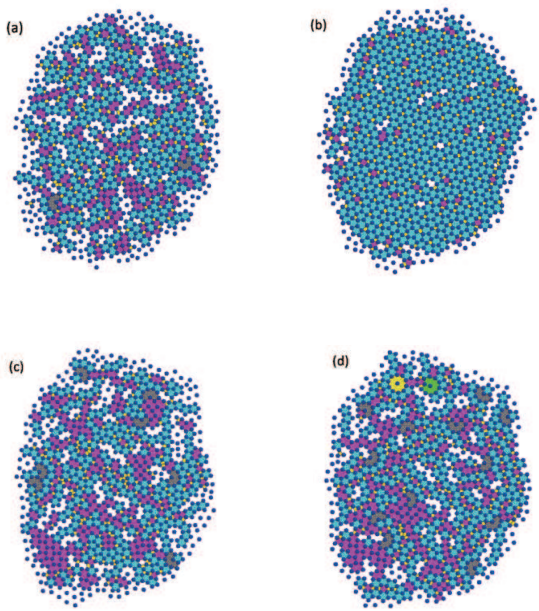}
%}
\caption{Evolution of the tiling structure.
Upper two panels are at $T^* = 0.3$ and $t/\tau = 3\times 10^2$(a) and
$10^5$(b).
Lower two panels are at $T^* = 0.15$ and $t/\tau = 3\times 10^2$(c) and
$10^5$(d).
}
\label{crystallization}       % Give a unique label
\end{figure*}%

\section{Discussion}
\label{discussion}
We have succeeded in vitrifying a simple monatomic liquid in two dimensions
by MD simulation and shown that the glassy state can live 
a fairly long time at low temperatures when the parameters of the 
LJG potential are chosen adequately. The Gaussian part of the LJG potential 
stabilizes the pentagonal configuration and packing of pentagons 
produces frustration in crystallization. Two competing length scales can have a 
great impact on the degree of disorder 
in the system. Also, in 3D systems\cite{hoang}, 
the glassy state can be formed if the LJG potential has appropriate parameters 
favoring the formation of an icosahedral local order. The effect of 
this frustration with packing is the origin of the stability of the glassy state.

The present simulation was carried out at zero pressure. In a previous 
report\cite{mizuguchi}, we investigated the glass transition of the same system at 
higher pressure corresponding to a constant denisity $\rho$, $\rho \pi r_0^2 = 1$. 
We obtained the temperature dependence of the energy while the system 
was heated at a constant rate, and showed that a transition
exists where the specific heat increases suddenly, which can be identified as the 
glass transition. A full report of the observation will be given in 
a forth coming paper\cite{mizuguchi2}.

The TTT diagram (Fig. \ref{TTTdiagram}) indicates that one can keep the glassy state 
of the LJG system for a fairly long time. Since we employ the free boundary 
condition in this study, the system can crystallize faster as compared with other 
confined boundary conditions. In fact, it was found that the amorphous structure of 
Lennard-Jones Ar confined in a nanoscale pore is more stable than a bulk 
Ar\cite{nishio}. Therefore, with other confined boundary conditions, our system will 
show stronger resistance against crystallization. Due to the 
stability of glass, this model provides a chance of further study about the glass 
transition and other related phenomena. In fact, our preliminary 
investigation\cite{mizuguchi} showed that the intermediate scattering function of 
the super cooled liquid state shows two step relaxation similar to those observed by 
experiments.

It is interesting to note that the LJG system can have crystalline states,
quasi-periodic sates and glassy states and the dynamics in these 
states, as well as  change in dynamics during the 
transformation, can be investigated and that the findings for the 
LJG system will serve as standards against which theroies can be tested. In 
particular, relation between the slow dynamics in glassy systems and the phason 
dynamics in quasi-crystals may be scrutinized.

\medskip
%\begin{acknowledgement}
\noindent
{\bf Acknowledgement}\\
We would like to thank Dr. Michael Engel of Stuttgart University
for providing us with the software on which most 
of the simulations and analysis for this work were carried out.
This work was supported in part by the Grant-in-Aid for Scientific Research
from the Ministry of Education, Culture, Sports, Science and Technology.
%\end{acknowledgement}
. 
%
% BibTeX users please use
% \bibliographystyle{}
% \bibliography{}

\begin{thebibliography}{99}
\bibitem{hiwatari}
B. Bernu, J. P. Hansen, Y. Hiwatari and G. Pastore,
Phys. Rev. A \textbf{36}, (1987) 4891.
\label{hiwatari}
%
\bibitem{kob}
%W. Kob and H.-C. Andersen, Phys. Rev. E\textbf{51}, (1995) 4626;
W. Kob and H.-C. Andersen, Phys. Rev. E \textbf{52}, (1995) 4134.%
\label{kob}%
%
\bibitem{mrs}
R. Bush, J. Schroers, and W. H.Wang, MRS Bulletin \textbf{32}, (2007) 620.
\label{mrs}
%
\bibitem{megen}
W. van Megen, Phys. Rev. E \textbf{76}, (2007) 061401.
\label{megen}
%
\bibitem{mct}
U. Bengtzelius, W. G\"{o}tze and A. Sj\"{o}lander, J. Phys. C {\bf 17}, (1984) 5915;
W. G\"{o}tze, in ``Liquids, Freezing and the Glass Transition" (edited by
J.~P.~Hansen, D.~Levesque and J.~Zinn-Justin), (North Holland 1989) p. 287; 
W. G\"otze, J. Phys.: Condens. Matt. {\bf 11}, (1999) A1.
\label{mct}
%4
\bibitem{fel}
T. Odagaki, Phys. Rev. Lett. {\bf 75}, (1995) 3701;
T. Odagaki, T, Yoshidome, A. Koyama, and A. Yoshimori,
J. Non-Crys. Solids, \textbf{352}, (2006) 4843;
T. Odagaki and A. Yoshimori, J. Non-Crys. Solids, to appear.
\label{fel}
%
\bibitem{replica}
M. M\'ezard, and G. Parisi, J. Chem. Phys. {\bf 111}, (1999) 1076.%
\label{replica}%
%
\bibitem{engel}
M. Engel and H.-R. Trebin, Phys. Rev. Lett. \textbf{98}, (2007) 225505;
M. Engel, Ph. D. Thesis (Stuttgart University, 2008).
\label{engel}
%
\bibitem{mizuguchi}
T. Mizuguchi, T. Odagaki, M. Umezaki, T.Koumyou, and J. Matsui,
\textit{Conference Proceedings 982 "Complex Systems"},
(AIP, New York 2008) 234.
\label{mizuguchi}
%
\bibitem{hoang}
Vo Van Hoang and T. Odagaki,
Physica B \textbf{403}, (2008) 3910
\label{hoang}
%
\bibitem{goukin}
Al-Lehyani, M. Widom, Y. Wang, N. Moghadam, G. M. Stocks and J. A. Moriaty,
Phys. Rev. B \textbf{64}, (2001) 075109.
\label{goukin}
%
\bibitem{rechtsman}
M. Rechtsman, F. Stillinger and S. Torquato, Phys. Rev. Lett. \textbf{95},
(2005) 228302; Phys. Rev. E \textbf{73}, (2006) 011406.
\label{rechtsman}
%
\bibitem{nishio}
K. Nishio, J. Koga, T. Yamaguchi and F. Yonezawa, 
J. Phys. Soc. Jpn \textbf{73}, (2004) 627.
\label{nishio}
%
\bibitem{TTT1}
H. Senapati, R. K. Kadiyala and C. A. Angell, 
J. Phys. Chem. \textbf{95}, (1991) 7050.
\label{TTT1}
%
\bibitem{TTT2}
S. Mukherjee, J. Schroers, W. L. Johnson and W.-K. Rhim, 
Phys. Rev. Lett. \textbf{94}, (2005) 245501.
\label{TTT2}
%
\bibitem{TTT_Fe}
A. V. Evteev, A. T. Kosilov, E. V. Levchenko and O. B. Logachev
J. Exp. Theor. Physics \textbf{101}, (2005) 521.
\label{TTT_Fe}
%
\bibitem{TTT_Ni}
Y. Zhang, Li Wang and W. Wang, 
J. Phys.:Condens. Matter \textbf{19}, (2007) 196106.
\label{TTT_Ni}
%
\bibitem{mizuguchi2}
T. Mizuguchi and T. Odagaki, in preparation.
\label{mizuguchi2}
%

%Author, \textit{Book title} (Publisher, place year) page numbers
% etc
\end{thebibliography}

\end{document}